\documentclass[journal,twocolumn,10pt,twoside]{IEEEtranTCOM}
\newtheorem{definition}{{Definition}}
\newtheorem{theorem}{{Theorem}}

\usepackage{cite}
\usepackage{amsmath,amssymb,amsfonts}
\usepackage{algorithmic}
\usepackage{graphicx}
\usepackage{textcomp}
\usepackage{xcolor}

\normalsize

\ifCLASSINFOpdf
\else
\fi

\begin{document}

\title{Constrained Gaussian Wasserstein Optimal Transport with Commutative Covariance Matrices}

\author{Jun~Chen, Jia~Wang, Ruibin~Li,  Han~Zhou, Wei~Dong,  Huan~Liu, and Yuanhao Yu
}

\maketitle

\begin{abstract}
	Optimal transport has found widespread applications in signal processing and machine learning. Among its many equivalent formulations, optimal transport seeks to reconstruct a random variable/vector with a prescribed distribution at the destination  while minimizing the expected distortion relative to a given random variable/vector at the source. However, in practice, certain constraints may render the optimal transport plan infeasible. In this work, we consider three types of constraints: rate constraints, dimension constraints, and channel constraints, motivated by perception-aware lossy compression, generative principal component analysis, and deep joint source-channel coding, respectively. Special attenion is given to the setting termed Gaussian Wasserstein optimal transport, where both the source and reconstruction variables are multivariate Gaussian, and the end-to-end distortion is measured by the mean squared error. We derive explicit results for the minimum achievable mean squared error under the three aforementioned constraints when the covariance matrices of the source and reconstruction variables commute. 
\end{abstract}


%
\IEEEpeerreviewmaketitle

\section{Introduction}

Since its inception, optimal transport has grown from a purely mathematical theory \cite{Villani03} into a powerful tool with widespread applications across numerous fields. Its influence extends so broadly that it is difficult to identify an area it has not impacted. In particular, optimal transport has had a profound effect on signal processing and machine learning, where it has shaped fundamental methodologies and inspired innovative approaches. This significant impact is well-documented in various survey papers \cite{KPTSR17,MMS25}, highlighting its role in advancing these domains.

Optimal transport admits many equivalent formulations. In this work, we adopt a formulation that frames the problem as reconstructing a random variable/vector 
$\hat{S}$ with a prescribed distribution 
$p_{\hat{S}}$ at the destination while minimizing the expected distortion relative to a given $S$ with distribution $p_S$
at the source. Mathematically, this corresponds to solving the following optimization problem:
\begin{align}
	\inf\limits_{p_{S\hat{S}}\in\Pi(p_S,p_{\hat{S}})}\mathbb{E}[c(S,\hat{S})],\label{eq:OT}
\end{align}
where $\Pi(p_S,p_{\hat{S}})$ denotes the set of joint distributions with marginals being $p_S$ and $p_{\hat{S}}$, and  $c(\cdot,\cdot)$ is the transport cost function, which in this formulation is more naturally interpreted as a distortion measure. In particular,  when $c(s,\hat{s})=\|s-\hat{s}\|^2$, the solution to 
\eqref{eq:OT} yields the squared Wassertein-2 distance between $p_S$ and $p_{\hat{S}}$, denoted by 
$W^2_2(p_S,p_{\hat{S}})$,

In practical applications, various constraints may render the optimal transport plan associated with the joint distribution $p_{S\hat{S}}$
that achieves the infimum in \eqref{eq:OT} infeasible. These constraints can arise from physical limitations, regulatory requirements, or structural restrictions imposed by the problem setting. Consequently, there is a need to study constrained optimal transport, which aims to develop methods for determining the best possible transport plan while ensuring compliance with the given constraints.

One such scenario occurs when the source and destination are connected by a rate-limited bit pipeline. In this case, continuous transport plans are no longer realizable, necessitating some form of discretization. This challenge has led to the investigation of distribution-preserving quantization \cite{LKK10, KZLK13} and output-constrained lossy source coding \cite{SLY15J1, SLY15J2}. 
Recently, this line of research has gained renewed interest due to the emergence of perception-aware lossy compression \cite{BM19, Matsumoto18,Matsumoto19,YWYML21, TW21, TA21, ZQCK21, CYWSGT22, SPCYK23, HWG24, SCKY24, SSK24, QSCKYSGT25, XLCZ24, XLCYZ24} and cross-domain lossy compression \cite{LZCK22, LZCK22J}. For related developments in the quantum setting, see \cite{GPC24}.

Another motivating scenario involves a dimensional bottleneck between the source and destination. In this case, it becomes essential to identify the minimum-dimensional representation of 
the source variable $S$ that enables faithful reconstruction. This concept underpins compressed sensing \cite{Donoho06, CRT06} and analog compression \cite{WV10}. In the lossy setting, it leads to techniques such as principal component analysis \cite{Jolliffe02}, among others. More generally, one may be interested in generative tasks where the reconstruction variable $\hat{S}$ 
does not need to be identical to the source variable $S$. This requires the development of dimension reduction methods specifically tailored for such purposes.

The emerging paradigm of generative communication \cite{QLCYW24}, which leverages deep generative models for joint source-channel coding \cite{BKG19, KG20, KG21, TG22, ETDG23}, also provides a  compelling impetus for studying constrained optimal transport. In this context, the source must communicate with the destination through a channel. Notably, unlike conventional communication problems, here the source-channel separation architecture  can be strictly suboptimal, even in the point-to-point scenario.

The present work focuses on the setting termed Gaussian Wasserstein optimal transport, where both the source and reconstruction variables are multivariate Gaussian, and the end-to-end distortion is measured by the mean squared error. Specifically, we assume that $S:=(S_1,S_2,\ldots,S_L)^T$ and $\hat{S}:=(\hat{S}_1,\hat{S}_2,\ldots,\hat{S}_L)^T$ are $L$-dimensional random vectors distributed according to $\mathcal{N}(\mu,\Sigma)$ and $\mathcal{N}(\hat{\mu},\hat{\Sigma})$, respectively, and $c(s,\hat{s})=\|s-\hat{s}\|^2$ for $s,\hat{s}\in\mathbb{R}^{L}$. Consequently, the solution to \eqref{eq:OT} is given by the squared Wasserstein-2 distance between $\mathcal{N}(\mu,\Sigma)$ and $\mathcal{N}(\hat{\mu},\hat{\Sigma})$ \cite{DL82,OP82,KS84,GS84} expressed as
\begin{align}
	&W^2_2(\mathcal{N}(\mu,\Sigma),\mathcal{N}(\hat{\mu},\hat{\Sigma}))\nonumber\\
	&=\|\mu-\hat{\mu}\|^2+\mathrm{tr}\left(\Sigma+\hat{\Sigma}-2(\Sigma^{\frac{1}{2}}\hat{\Sigma}\Sigma^{\frac{1}{2}})^{\frac{1}{2}}\right).\label{eq:GW}
\end{align}
Special attention is given to the case where the covariance matrices $\Sigma$ and $\hat{\Sigma}$ are positive definite and commute. This allows us to write them as $\Sigma=\Theta\Lambda\Theta^T$ and $\hat{\Sigma}=\Theta\hat{\Lambda}\Theta^T$, where $\Theta$ is a unitary matrix, while $\Lambda:=\mathrm{diag}(\lambda_1,\lambda_2,\ldots,\lambda_L)$ and $\hat{\Lambda}:=\mathrm{diag}(\hat{\lambda}_1,\hat{\lambda}_2,\ldots,\hat{\lambda}_L)$ are diagonal matrices with positive diagonal entries. For this case, \eqref{eq:GW} simplifies to
\begin{align}
	&W^2_2(\mathcal{N}(\mu,\Sigma),\mathcal{N}(\hat{\mu},\hat{\Sigma}))\nonumber\\
	&=\|\mu-\hat{\mu}\|^2+\sum\limits_{\ell=1}^L\left(\sqrt{\lambda_{\ell}}-\sqrt{\hat{\lambda}_{\ell}}\right)^2.\label{eq:GWscalar}
\end{align}
The optimal transport plan that achieves \eqref{eq:GWscalar} is given by
\begin{align}
	\hat{S}=\Theta \mathrm{diag}\left(\sqrt{\frac{\hat{\lambda}_1}{\lambda_1}},\sqrt{\frac{\hat{\lambda}_2}{\lambda_2}},\ldots,\sqrt{\frac{\hat{\lambda}_L}{\lambda_L}}\right)\Theta^T (S-\mu)+\hat{\mu},
\end{align}
which is an affine transformation. 
For notational simplicity,  we henceforth assume $\mu=\hat{\mu}=0$, $\Theta=I$ (i.e., $\Sigma=\Lambda$ and $\hat{\Sigma}=\hat{\Lambda}$), and
\begin{align} \lambda_{1}\hat{\lambda}_1\geq\lambda_2\hat{\lambda}_2\geq\ldots\geq\lambda_{L}\hat{\lambda}_L.
\end{align} 
It will be seen that the Gaussian Wasserstein optimal transport problem serves as an ideal framework for examining the three key constraints discussed earlier: rate constraints, dimension constraints, and channel constraints. To ensure a coherent treatment, we adopt the asymptotic setting rather than the one-shot setting where 
 a single reconstruction variable/vector $\hat{S}$ is generated for a single source variable/vector $S$. Specifically, we consider the task of generating an i.i.d. reconstruction sequence $\hat{S}^n$ with $\hat{S}(t)\sim\mathcal{N}(0,\hat{\Lambda})$,  $t=1,2,\ldots,n$, in response to an i.i.d. source sequence $S^n$ with $S(t)\sim\mathcal{N}(0,\Lambda)$,   $t=1,2,\ldots,n$, while minimizing the average distortion 
\begin{align}
	\frac{1}{n}\sum\limits_{t=1}^n\mathbb{E}[\|S(t)-\hat{S}(t)\|^2].
\end{align}
In the absence of constraints, there is no fundamental difference between the one-shot and asymptotic settings, as transport can be performed in a symbol-by-symbol manner without loss of optimality.



Our main contributions are as follows:
\begin{enumerate}
	\item For rate-constrained optimal transport, we distinguish between the case with unlimited common randomness and the case with no common randomness, deriving reverse waterfilling-type formulas for the minimum achievable distortion in both cases.
	
	\item For dimension-constrained optimal transport, we extend principal component analysis to generative tasks. 
	
	\item For channel-constrained optimal transport, we provide a systematic comparison of separation-based, uncoded, and hybrid schemes.
\end{enumerate}

The remainder of this paper is organized as follows. Sections \ref{sec:rate}, \ref{sec:dimension}, and \ref{sec:channel} explore rate-constrained, dimension-constrained, and channel-constrained optimal transport, respectively. Recurring themes,  important connections, and key differences across the three types of constraints are highlighted throughout these sections. Finally, we conclude the paper in Section \ref{sec:conclusion}.

Throughout this paper, we adopt the standard notation for information measures: $I(\cdot;\cdot)$ for mutual information and $h(\cdot)$ for differential entropy.
The set of nonnegative numbers is denoted by
 $\mathbb{R}_+$. We define $\log^+(a):=\max\{\log(a),0\}$, $(a)_+:=\max\{x,0\}$, and $a\wedge b:=\min\{a,b\}$. A Gaussian distribution with mean $\mu$ and covariance matrix $\Sigma$ is represented as $\mathcal{N}(\mu,\Sigma)$. For brevity, we use $X^n$ to denote the sequence $\{X(t)\}_{t=1}^n$.  Summations of the form
$\sum_{i=j}^ka_i$ are defined to be zero whenever $j>k$. The expectation, trace, floor, and ceiling operators are denoted by
 $\mathbb{E}[\cdot]$, $\mathrm{tr}(\cdot)$, $\lfloor\cdot\rfloor$, and $\lceil\cdot\rceil$, respectively.
For two matrices $A$ and $B$,  the notation $A\preceq B$ indicates that $B-A$ is positive semidefinite.
Finally, we use $\log$ and $\ln$ to denote  logarithms with bases $2$ and $e$, respectively.

\section{Rate-Constrained Optimal Transport}\label{sec:rate}

In this section, we examine the scenario where the source and destination are connected by a rate-limited bit pipeline, necessitating the deployment of an encoder and a decoder. Given the source sequence $S^n$, the encoder produces a length-$m$ bit string $B^m\in\{0,1\}^m$
and transmits it to the destination via the bit pipeline. Upon receiving the bit string, the decoder generates a reconstruction sequence
$\hat{S}^n$ with the prescribed distribution while minimizing the end-to-end distortion. This scenario is first studied in \cite{SLY15J2} from an information-theoretic perspective. By focusing on the Gaussian Wasserstein  setting, we are able to obtain  explicit reverse waterfilling-type results by leveraging convex optimization techniques.
Notably, unlike conventional source coding problems, the minimum achievable distortion in this setting depends on the availability of common randomness. Accordingly, we structure our analysis to account for this dependency.


\subsection{Unlimited Common Randomness}

 Here, the encoder and decoder are assumed to share a random seed $Q$. Accordingly, their operations are governed by the conditional distributions $p_{B^m|S^nQ}$ and $p_{\hat{S}^n|B^mQ}$, respectively. The overall system is characterized by the joint distribution $p_{S^nQB^m\hat{S}^n}$ factorized as
	\begin{align}
		p_{S^nQB^m\hat{S}^n}=p_{S^n}p_Qp_{B^m|S^nQ}p_{\hat{S}^n|B^mQ},\label{eq:factor3}
	\end{align}
where $p_{S^n}=p^n_S$ with $p_S=\mathcal{N}(0,\Lambda)$.

\begin{definition}
	With common randomness, a distortion level $D$ is said to be achievable under a rate constraint $R$ if, for  all sufficiently large $n$, there exist a seed distribution $p_Q$, an encoding distribution $p_{B^m|S^nQ}$, and a decoding distribution $p_{\hat{S}^n|B^mQ}$ such that  
	\begin{align}
			&\frac{m}{n}\leq R,\\
		&\frac{1}{n}\sum\limits_{t=1}^n\mathbb{E}[\|S(t)-\hat{S}(t)\|^2]\leq D,
	\end{align}
and the reconstruction sequence $\hat{S}^n$ follows the i.i.d. distribution $p_{\hat{S}^n}=p^n_{\hat{S}}$ with $p_{\hat{S}}=\mathcal{N}(0,\hat{\Lambda})$.
	The infimum of all achievable distortion levels $D$ under the rate constraint $R$ with common randomness is denoted by $\underline{D}_r(R)$.
\end{definition}

The following result provides an explicit characterization of $\underline{D}_r(R)$. Its proof can be found in Appendix \ref{app:ratecr}.

\begin{theorem}\label{thm:ratecr}
For $R\geq 0$,
\begin{align}
\underline{D}_r(R)=\sum\limits_{\ell=1}^L\left(\lambda_{\ell}+\hat{\lambda}_{\ell}-2\sqrt{(1-2^{-2\underline{R}_{\ell}(R)})\lambda_{\ell}\hat{\lambda}_{\ell}}\right),
\end{align}
where
\begin{align}
	\underline{R}_{\ell}(R):=
	\frac{1}{2}\log\left(\frac{1+\sqrt{1+\alpha\lambda_{\ell}\hat{\lambda}_{\ell}}}{2}\right),\quad\ell=1,2,\ldots,L,\label{eq:rate1}
\end{align}
with $\alpha$ being the unique nonnegative number satisfying 
\begin{align}
	\frac{1}{2}\sum\limits_{\ell=1}^L\log\left(\frac{1+\sqrt{1+\alpha\lambda_{\ell}\hat{\lambda}_{\ell}}}{2}\right)=R.
\end{align}

\end{theorem}

Theorem \ref{thm:ratecr} admits a natural operational interpretation. With the availability of common randomness,  as $n\rightarrow\infty$, it takes $R_{\ell}:=I(S_{\ell},\hat{S}_{\ell})$ bits per symbol to simulate $\hat{S}^n_{\ell}$ such that  its pairwise distributions with $S^n_{\ell}$, i.e.,  $p_{S_{\ell}(t)\hat{S}_{\ell}(t)}$, $t=1,2,\ldots,n$,  are all equal to a prescribed  bivariate Gaussian  distribution $p_{S_{\ell}\hat{S}_{\ell}}$, where $S_{\ell}\sim\mathcal{N}(0,\lambda_{\ell})$ and $\hat{S}_{\ell}\sim\mathcal{N}(0,\hat{\lambda}_{\ell})$ for $\ell=1,2,\ldots,L$.  Assuming the correlation coefficient of $S_{\ell}$ and $\hat{S}_{\ell}$ is $\rho_{\ell}\geq 0$,
we have
\begin{align}
	R_{\ell}=\frac{1}{2}\log\left(\frac{1}{1-\rho^2_{\ell}}\right),
\end{align}
which implies
\begin{align}
	\rho_{\ell}=\sqrt{1-2^{-2R_{\ell}}},\quad\ell=1,2,\ldots,L.
\end{align}
Consequently,
\begin{align}
	&\frac{1}{n}\sum\limits_{t=1}^n\mathbb{E}[\|S(t)-\hat{S}(t)\|^2]\nonumber\\
	&=\frac{1}{nL}\sum\limits_{t=1}^n\sum\limits_{\ell=1}^n\mathbb{E}[(S_{\ell}(t)-\hat{S}_{\ell}(t))^2]\nonumber\\
	&=\sum\limits_{\ell=1}^L\mathbb{E}[(S_{\ell}-\hat{S}_{\ell})^2]\nonumber\\
	&=\sum\limits_{\ell=1}^L\left(\lambda_{\ell}+\hat{\lambda}_{\ell}-2\sqrt{(1-2^{-2R_{\ell}})\lambda_{\ell}\hat{\lambda}_{\ell}}\right).
\end{align}
This leads to the following rate allocation problem:
\begin{align}
	&\hspace{-0.1in}\min\limits_{(R_1,R_2\ldots,R_L)\in\mathbb{R}^L_+}\sum\limits_{\ell=1}^L\left(\lambda_{\ell}+\hat{\lambda}_{\ell}-2\sqrt{(1-2^{-2R_{\ell}})\lambda_{\ell}\hat{\lambda}_{\ell}}\right)\\
	&\hspace{-0.1in}\mbox{s.t.}\quad \sum\limits_{\ell=1}^LR_{\ell}\leq R,
\end{align}
with the minimizer  given by  \eqref{eq:rate1}.

\subsection{No Common Randomness}

 Here, the encoder and decoder are assumed to operate without a shared random seed. Specifically, their operations are governed by  the conditional distributions $p_{B^m|S^n}$ and $p_{\hat{S}^n|B^m}$, respectively. The overall system is characterized by the joint distribution $p_{S^nB^m\hat{S}^n}$ factorized as
\begin{align}
	p_{S^nB^m\hat{S}^n}=p_{S^n}p_{B^m|S^n}p_{\hat{S}^n|B^m},\label{eq:factor4}
\end{align}
where $p_{S^n}=p^n_S$ with $p_S=\mathcal{N}(0,\Lambda)$.

\begin{definition}
	Without common randomness, a distortion level $D$ is said to be achievable under a rate constraint $R$ if, for all sufficiently large $n$, there exist  an encoding distribution $p_{B^m|S^n}$ and a decoding distribution $p_{\hat{S}^n|B^m}$ such that 
	\begin{align}
		&\frac{m}{n}\leq R,\\
		&\frac{1}{n}\sum\limits_{t=1}^n\mathbb{E}[\|S(t)-\hat{S}(t)\|^2]\leq D,
	\end{align}
and the reconstruction sequence $\hat{S}^n$ follows the i.i.d. distribution $p_{\hat{S}^n}=p^n_{\hat{S}}$ with $p_{\hat{S}}=\mathcal{N}(0,\hat{\Lambda})$.
The infimum of all achievable distortion levels $D$ under the rate constraint $R$ without common randomness is denoted by $\overline{D}_r(R)$.
\end{definition}

The following result provides an explicit characterization of $\overline{D}_r(R)$. Its proof can be found in Appendix \ref{app:ratencr}.

\begin{theorem}\label{thm:ratencr}
For $R\geq 0$,
\begin{align}
	\overline{D}_r(R)=\sum\limits_{\ell=1}^L\left(\lambda_{\ell}+\hat{\lambda}_{\ell}-2(1-2^{-2\overline{R}_{\ell}(R)})\sqrt{\lambda_{\ell}\hat{\lambda}_{\ell}}\right),
\end{align}
where
\begin{align}
	\overline{R}_{\ell}(R):=
		\frac{1}{2}\log^+\left(\frac{\sqrt{\lambda_{\ell}\hat{\lambda}_{\ell}}}{\beta}\right),\quad\ell=1,2,\ldots,L,\label{eq:rate2}
\end{align}
with $\beta$ being the unique number in $(0,\sqrt{\lambda_1\hat{\lambda}_1}]$ satisfying
\begin{align}
	\frac{1}{2}\sum\limits_{\ell=1}^L\log^+\left(\frac{\sqrt{\lambda_{\ell}\hat{\lambda}_{\ell}}}{\beta}\right)=R.
\end{align} 
\end{theorem}

Theorem \ref{thm:ratencr} also admits a natural operational interpretation. First, the encoder maps $S^n_{\ell}$ to $U^n_{\ell}$  using a rate-$R_{\ell}$ vector quantizer such that
\begin{align}
	\frac{1}{n}\sum\limits_{t=1}^n\mathbb{E}[(S_{\ell}(t)-U_{\ell}(t))^2]= \lambda_{\ell}-\gamma_{\ell}\approx 2^{-2R_{\ell}}\lambda_{\ell},
\end{align}
where  $\gamma_{\ell}:=\frac{1}{n}\sum_{t=1}^n\mathbb{E}[U^2_{\ell}(t)]$ for $\ell=1,2,\ldots,L$. Next, it converts $U^n_{\ell}$ to $\hat{U}^n_{\ell}(t)$ via the transport plan induced by a coupling satisfying
\begin{align}
	\frac{1}{n}\sum\limits_{t=1}^n\mathbb{E}[(U_{\ell}(t)-\hat{U}_{\ell}(t))^2]\approx(\sqrt{\gamma}_{\ell}-\sqrt{\hat{\gamma}}_{\ell})^2,
\end{align}
where $\hat{U}^n_{\ell}$ is distributed according to the output of a rate-$R_{\ell}$ quantizer with the reconstruction $\hat{S}^n_{\ell}$ serving as a fictitious source, and $\hat{\gamma}_{\ell}:=\frac{1}{n}\sum_{t=1}^n\mathbb{E}[\hat{U}^2_{\ell}(t)]$ for $\ell=1,2,\ldots,L$.  The decoder then generates $\hat{S}^n_{\ell}$  from  $\hat{U}^n_{\ell}$ by performing dequantization (also known as posterior sampling), yielding
\begin{align}
	\frac{1}{n}\sum\limits_{t=1}^n\mathbb{E}[(\hat{S}_{\ell}(t)-\hat{U}_{\ell}(t))^2]=\hat{\lambda}_{\ell}-\hat{\gamma}_{\ell}\approx 2^{-2R_{\ell}}\hat{\lambda}_{\ell},
\end{align}
for $\ell=1,2,\ldots,L$. Consequently,
\begin{align}
	&\frac{1}{n}\sum\limits_{t=1}^n\mathbb{E}[\|S(t)-\hat{S}(t)\|^2]\nonumber\\
	&=\frac{1}{n}\sum\limits_{t=1}^n\sum\limits_{\ell=1}^L\mathbb{E}[(S_{\ell}(t)-U_{\ell}(t))^2+(U_{\ell}(t)-\hat{U}_{\ell}(t))^2\nonumber\\
	&\hspace{2in}+(\hat{S}_{\ell}(t)-\hat{U}(t))^2]\nonumber\\
	&=\sum\limits_{\ell=1}^L(\lambda_{\ell}+\hat{\lambda}_{\ell}-2\sqrt{\gamma_{\ell}\hat{\gamma}_{\ell}})\nonumber\\
	&\approx\sum\limits_{\ell=1}^L\left(\lambda_{\ell}+\hat{\lambda}_{\ell}-2(1-2^{-2R_{\ell}})\sqrt{\lambda_{\ell}\hat{\lambda}_{\ell}}\right).
\end{align}
This leads to the following rate allocation problem:
\begin{align}
	&\hspace{-0.1in}\min\limits_{(R_1,R_2\ldots,R_L)\in\mathbb{R}^L_+}\sum\limits_{\ell=1}^L\left(\lambda_{\ell}+\hat{\lambda}_{\ell}-2(1-2^{-2R_{\ell}})\sqrt{\lambda_{\ell}\hat{\lambda}_{\ell}}\right)\\
	&\hspace{-0.1in}\mbox{s.t.}\quad \sum\limits_{\ell=1}^LR_{\ell}\leq R,
\end{align}
with the minimizer given by  \eqref{eq:rate2}.

\begin{figure}[htbp]
	\centerline{\includegraphics[width=7cm]{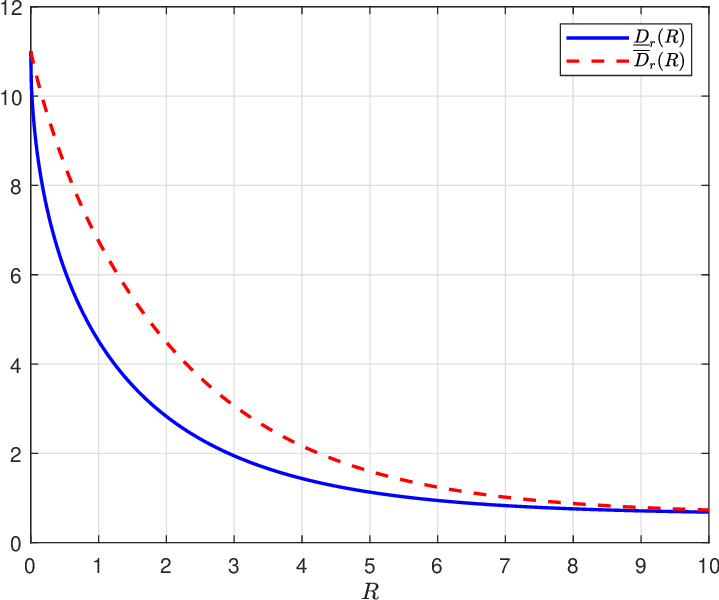}} \caption{Plots of $\underline{D}_r(R)$ and $\overline{D}_r(R)$ for the case where $(\lambda_1,\lambda_2,\lambda_3)=(2,3,1)$ and $(\hat{\lambda}_1,\hat{\lambda}_2,\hat{\lambda}_3)=(3,1,1)$.}
	\label{fig:rate} 
\end{figure}

\begin{table}[h]
	\centering
	\begin{tabular}{|c|c|c|c|}
		\hline
		 & $R=0.1$ & $R=2.1$ & $R=4.1$\\
		 \hline
		 $\underline{R}_1(R)$ & 0.058 & 0.929 &  1.641 \\ \hline
		$\underline{R}_2(R)$  &  0.031  & 0.726 & 1.407  \\  \hline
		$\underline{R}_3(R)$ & 0.011 & 0.445 &  1.051\\ \hline
		$\overline{R}_1(R)$ & 0.1 & 0.999  & 1.665 \\ \hline
		$\overline{R}_2(R)$ & 0 &  0.749  & 1.415 \\  \hline
		$\overline{R}_3(R)$ & 0 & 0.353  & 1.019\\ \hline
	\end{tabular}
	\caption{Comparison of $(\underline{R}_1(R),\underline{R}_2(R),\underline{R}_3(R))$ and $(\overline{R}_1(R),\overline{R}_2(R),\overline{R}_3(R))$ for the case where $(\lambda_1,\lambda_2,\lambda_3)=(2,3,1)$ and $(\hat{\lambda}_1,\hat{\lambda}_2,\hat{\lambda}_3)=(3,1,1)$.}
	\label{tab:rate}
\end{table}

It is instructive to compare $\underline{D}_r(R)$ and $\overline{D}_r(R)$,
as well as their associated rate allocation schemes (see also Fig. \ref{fig:rate} and Table \ref{tab:rate}). Clearly, $\underline{D}_r(R)<\overline{D}_r(R)$ for all $R>0$. Both $\underline{D}_r(R)$ and $\overline{D}_r(R)$ approach 
\begin{align}
	D_{\max}:=\sum\limits_{\ell=1}^L(\lambda_{\ell}+\hat{\lambda}_{\ell})
\end{align}
as $R\rightarrow 0$ and approach
\begin{align}
	D_{\min}:=\sum\limits_{\ell=1}^L\left(\sqrt{\lambda}_{\ell}-\sqrt{\hat{\lambda}_{\ell}}\right)^2
\end{align}
as $R\rightarrow\infty$, where $D_{\max}$ and $D_{\min}$ are, respectively, the distortion achieved by generating $\hat{S}^n$ independently of $S^n$ 
and the distortion achieved by unconstrained optimal transport (cf. \eqref{eq:GWscalar}). For each $\ell=1,2,\ldots,L$, both
$\underline{R}_{\ell}(R)$ and $\overline{R}_{\ell}(R)$ are increasing functions of $R$. For a fixed $R$, the ordering of $\underline{R}_{1}(R), \underline{R}_{2}(R),\ldots,\underline{R}_{L}(R)$ is determined by that of $\lambda_{1}\hat{\lambda}_1,\lambda_{2}\hat{\lambda}_2,\ldots,\lambda_{L}\hat{\lambda}_L$, with larger values of  $\lambda_{\ell}\hat{\lambda}_{\ell}$ corresponding to higher $\underline{R}_{\ell}(R)$. The same ordering applies to  $\overline{R}_{1}(R),\overline{R}_{2}(R),\ldots,\overline{R}_{L}(R)$. It will be seen that the value of $\lambda_{\ell}\hat{\lambda}_{\ell}$  as a measure of significance is a recurring theme across other constrained optimal transport problems. On the other hand, the two rate allocation schemes also have some notable differences. In particular, $\underline{R}_{1}(R), \underline{R}_{2}(R),\ldots,\underline{R}_{L}(R)$ are strictly positive whenever $R>0$, whereas some of $\overline{R}_{1}(R), \overline{R}_{2}(R),\ldots,\overline{R}_{L}(R)$ can be zero when $R$ is sufficiently small. Moreover, for a given total rate $R>0$, the individual rates $\overline{R}_{1}(R), \overline{R}_{2}(R),\ldots,\overline{R}_{L}(R)$
exhibit greater variation across all components compared to their counterparts $\underline{R}_{1}(R), \underline{R}_{2}(R),\ldots,\underline{R}_{L}(R)$.

\section{Dimension-Constrained Optimal Transport}\label{sec:dimension}

In this section, we consider the scenario where the bottleneck between the source and destination takes the form of a dimensionality constraint. As a result, the encoder must identify a low-dimensional representation of the source sequence $S^n$, from which the decoder generates a reconstruction sequence $\hat{S}^n$, ensuring the prescribed distribution while minimizing the end-to-end distortion.

It is clear that a certain regularity condition needs to be imposed on the encoder since otherwise the source sequence could be losslessly represented using a single real number. For this reason, we require the encoder to be a linear mapping $\phi$. On the other hand, the decoder is allowed to be a stochastic function governed by the conditional distribution $p_{\hat{S}^n|\phi(S^n)}$. The overall system is characterized by the joint distribution $p_{S^n\phi(S^n)\hat{S}^n}$ factorized as
\begin{align}
	p_{S^n\phi(S^n)\hat{S}^n}=p_{S^n}p_{\phi(S^n)|S^n}p_{\hat{S}^n|\phi(S^n)},
\end{align}
where $p_{S^n}=p^n_S$ with $p_S=\mathcal{N}(0,\Lambda)$.

\begin{definition}
	A distortion level $D$ is said to be achievable under a normalized dimension constraint $\Gamma$ if, for all sufficiently large $n$, there exist  a linear encoding function $\phi:\mathbb{R}^{nL}\rightarrow\mathbb{R}^m$ and a decoding distribution $p_{\hat{S}^n|\phi(S^n)}$ such that 
	\begin{align}
		&\frac{m}{n}\leq \Gamma,\\
		&\frac{1}{n}\sum\limits_{t=1}^n\mathbb{E}[\|S(t)-\hat{S}(t)\|^2]\leq D,
	\end{align}
	and the reconstruction sequence $\hat{S}^n$ follows the i.i.d. distribution $p_{\hat{S}^n}=p^n_{\hat{S}}$ with $p_{\hat{S}}=\mathcal{N}(0,\hat{\Lambda})$.
	The infimum of all achievable distortion levels $D$ under the normalized dimension constraint $\Gamma$  is denoted by $D_d(\Gamma)$.
\end{definition}

We first prove the following one-shot result. Its proof can be found in Appendix \ref{app:dimension}.

\begin{theorem}\label{thm:dimension}
Given $S\sim\mathcal{N}(0,\Lambda)$, for any linear encoding function $\phi:\mathbb{R}^L\rightarrow\mathbb{R}^K$  and decoding distribution $p_{\hat{S}|\phi(S)}$ such that the induced distribution $p_{\hat{S}}=\mathcal{N}(0,\hat{\Lambda})$,  we have
\begin{align}
\mathbb{E}[\|S-\hat{S}\|^2]\geq\sum\limits_{\ell=1}^{K\wedge L}\left(\sqrt{\lambda_{\ell}}-\sqrt{\hat{\lambda}_{\ell}}\right)^2+\sum\limits_{\ell=(K\wedge L)+1}^L(\lambda_{\ell}+\hat{\lambda}_{\ell}).
\end{align}
Moreover, this lower bound is achieved by selecting the first $K\wedge L$ components  of $S$, scaling them to obtain $\hat{S}_{1}, \hat{S}_{2}, \ldots,\hat{S}_{K\wedge L}$, and generating the remaining components of $\hat{S}$ from scratch. 
\end{theorem}

The scheme achieving the lower bound in Theorem \ref{thm:dimension} can be interpreted as a generative variant of principal component analysis, where the selection rule  is determined by the ordering of $\lambda_1\hat{\lambda}_1,\lambda_2\hat{\lambda}_2,\ldots,\lambda_L\hat{\lambda}_L$. This selection rule simplifies to that of conventional principal component analysis \cite{Jolliffe02}
when $\Lambda=\hat{\Lambda}$.

Since $S^n$ and $\hat{S}^n$ can be regarded as Gaussian random vectors of dimension $nL$, with their covariance matrices preserving a diagonal structure, 
 we can directly infer the following result from Theorem \ref{thm:dimension}.

\begin{theorem}\label{thm:dimension_asymptotic}
	For $\Gamma\in[0,L]$,
	\begin{align}
	D_d(\Gamma)=&\sum\limits_{\ell=1}^{\lfloor\Gamma\rfloor}\left(\sqrt{\lambda_{\ell}}-\sqrt{\hat{\lambda}_{\ell}}\right)^2\nonumber\\
	&+(\Gamma-\lfloor\Gamma\rfloor)\left(\sqrt{\lambda_{\lceil\Gamma\rceil}}-\sqrt{\hat{\lambda}_{\lceil\Gamma\rceil}}\right)^2\nonumber\\
	&+(\lceil\Gamma\rceil-\Gamma)(\lambda_{\lceil\Gamma\rceil}+\hat{\lambda}_{\lceil\Gamma\rceil})+\sum\limits_{\ell=\lceil\Gamma\rceil+1}^L(\lambda_{\ell}+\hat{\lambda}_{\ell}).
	\end{align}
Moreover, $D_d(\Gamma)=D_d(L)$ for $\Gamma>L$.
\end{theorem}

\begin{figure}[htbp]
	\centerline{\includegraphics[width=7cm]{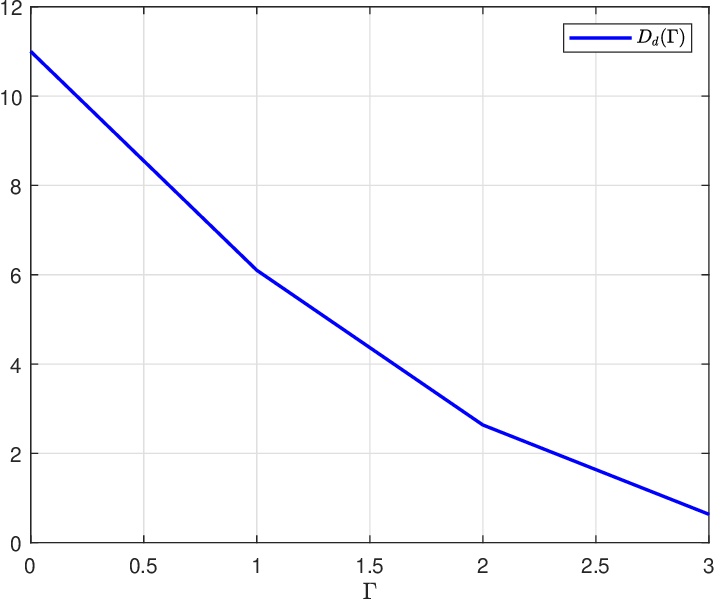}} \caption{Plot of $D_d(\Gamma)$  for the case where $(\lambda_1,\lambda_2,\lambda_3)=(2,3,1)$ and $(\hat{\lambda}_1,\hat{\lambda}_2,\hat{\lambda}_3)=(3,1,1)$.}
	\label{fig:dimension} 
\end{figure}

$D_d(\Gamma)$ is a decreasing, convex, piecewise linear function of $\Gamma$, approaching $D_{\max}$ as $\Gamma\rightarrow 0$ and $D_{\min}$ as $\Gamma\rightarrow L$ (see Fig. \ref{fig:dimension}). In comparison,  $\underline{D}_r(R)$ and $\overline{D}_r(R)$ exhibit similar overall behavior but are strictly convex.


\section{Channel-Constrained Optimal Transport}\label{sec:channel}

In this section, we consider the scenario where the transport must pass through 
an additive unit-variance white Gaussian noise channel $p_{Y|X}$, denoted by $\mathrm{AWGN}(1)$, where $Y=X+N$ with $N\sim\mathcal{N}(0,1)$ independent of $X$. This scenario is first studied in \cite[Section V]{QLCYW24} for  the degenerate case $\Lambda=\hat{\Lambda}$.
We will again distinguish between the cases with and without common randomness.

With common randomness, the encoder and decoder are assumed to share a random seed $Q$. Accordingly, their operations are governed by the conditional distributions $p_{X^n|S^nQ}$ and $p_{\hat{S}^n|Y^nQ}$, respectively. The overall system is characterized by the joint distribution $p_{S^nQX^nY^n\hat{S}^n}$ factorized as
 \begin{align}
 	p_{S^nQX^nY^n\hat{S}^n}=p_{S^n}p_Qp_{X^n|S^nQ}p_{Y^n|X^n}p_{\hat{S}^n|Y^nQ},\
 \end{align}
 where $p_{S^n}=p^n_S$ with $p_S=\mathcal{N}(0,\Lambda)$ and $p_{Y^n|X^n}=p^n_{Y|X}$ with $p_{Y|X}=\mathrm{AWGN}(1)$.


\begin{definition}\label{def:OTCw}
	With common randomness, a distortion level $D$ is said to be achievable through $\mathrm{AWGN}(1)$ under an input power constraint $P$ if, for all sufficiently large $n$, there exist a seed distribution $p_Q$, an encoding distribution $p_{X^n|S^nQ}$, and a decoding distribution $p_{\hat{S}^n|Y^nQ}$ such that 
	\begin{align}
		&\frac{1}{n}\sum\limits_{t=1}^n\mathbb{E}[X^2(t)]\leq P,\label{eq:constraint2}\\
		&\frac{1}{n}\sum\limits_{t=1}^n\mathbb{E}[\|S(t)-\hat{S}(t)\|^2]\leq D,\label{eq:constraint3}		
	\end{align}
	and the reconstruction sequence $\hat{S}^n$ follows the i.i.d. distribution $p_{\hat{S}^n}=p^n_{\hat{S}}$ with $p_{\hat{S}}=\mathcal{N}(0,\hat{\Lambda})$. The infimum of all achievable distortion levels $D$ through $\mathrm{AWGN}(1)$ under the input power constraint $P$ with commom randomness is denoted by $\underline{D}_c(P)$.
\end{definition}

According to \cite[Theorem 1]{QLCYW24}, when unlimited common randomness is available,
 the source-channel separation theorem holds for channel-constrained optimal transport,  namely, there is no loss of optimality in first converting the channel to a bit pipeline using error correction codes and then performing rate-constrained optimal transport. Combining this result with the capacity formula of $\mathrm{AWGN}(1)$, we obtain
\begin{align}
	\underline{D}_c(P)=\underline{D}_r\left(\frac{1}{2}\log(P+1)\right).
\end{align}

Without common randomess,  the encoding and decoding operations are governed by  the conditional distributions $p_{X^n|S^n}$ and $p_{\hat{S}^n|Y^n}$, respectively. The overall system is characterized by the joint distribution $p_{S^nX^nY^n\hat{S}^n}$ factorized as
\begin{align}
	p_{S^nX^nY^n\hat{S}^n}=p_{S^n}p_{X^n|S^n}p_{Y^n|X^n}p_{\hat{S}^n|Y^n},\
\end{align}
where $p_{S^n}=p^n_S$ with $p_S=\mathcal{N}(0,\Lambda)$ and $p_{Y^n|X^n}=p^n_{Y|X}$ with $p_{Y|X}=\mathrm{AWGN}(1)$.

\begin{definition}\label{def:OTCw/o}
	Without common randomness, a distortion level $D$ is said to be achievable through $\mathrm{AWGN}(1)$ under an input power constraint $P$ if, for all sufficiently large $n$, there exist an encoding distribution $p_{X^n|S^n}$ and a decoding distribution $p_{\hat{S}^n|Y^n}$ such that 
	\begin{align}
		&\frac{1}{n}\sum\limits_{t=1}^n\mathbb{E}[X^2(t)]\leq P,\label{eq:power}\\
		&\frac{1}{n}\sum\limits_{t=1}^n\mathbb{E}[\|S(t)-\hat{S}(t)\|^2]\leq D,		
	\end{align}
	and the reconstruction sequence $\hat{S}^n$ follows the i.i.d. distribution $p_{\hat{S}^n}=p^n_{\hat{S}}$ with $p_{\hat{S}}=\mathcal{N}(0,\hat{\Lambda})$. The infimum of all achievable distortion levels $D$ through $\mathrm{AWGN}(1)$ under the input power constraint $P$ without commom randomness is denoted by $\overline{D}_c(P)$.
\end{definition}

Let $D^{(s)}_c(P)$ denote the minimum achievable distortion under the separation-based scheme, i.e., 
\begin{align}
D^{(s)}_c(P):=\overline{D}_r\left(\frac{1}{2}\log(P+1)\right).\label{eq:Dsep}
\end{align}
It turns out that $D^{(s)}_c(P)$ is just an upper bound on $\overline{D}_c(P)$. As we will demonstrate, the source-channel separation architecture is generally suboptimal for channel-constrained optimal transport when no common randomness is available. 

To this end, consider the following uncoded scheme. The encoder transmits $X^n:=\sqrt{\frac{P}{\lambda_{1}}}S^n_{1}$, obtained by scaling the first component of each source symbol to meet the power contraint while discarding the other components $S^n_2,S^n_3,\ldots,S^n_L$; given the channel output $Y^n$, the decoder sets  $\hat{S}^n_{1}:=\sqrt{\frac{\hat{\lambda}_{1}}{P+1}}Y^n$ and generates the remaining components $\hat{S}^n_2,\hat{S}^n_3,\ldots,\hat{S}^n_L$ of the reconstruction sequence  from scratch. It can be verified that the resulting distortion is given by
\begin{align}
	D^{(u)}_c(P):=-2\sqrt{\frac{P}{P+1}\lambda_{1}\hat{\lambda}_{1}}+\sum\limits_{\ell=1}^L(\lambda_{\ell}+\hat{\lambda}_{\ell}).
\end{align}

The following result, which is a ``noisy" variant of Theorem \ref{thm:dimension_asymptotic} for the special case $\Gamma=1$ and a generalization of the one-shot optimality result \cite[Theorem 3]{QLCYW24} for the degenerate case $\Lambda=\hat{\Lambda}$,  
 indicates that this uncoded scheme is the best one among all linear schemes. Its proof can be found in Appendix \ref{app:linear}.

\begin{theorem}\label{thm:linear}
	Let $X^n:=\phi(S^n)$ be the channel input induced by a linear mapping $\phi$
	satisfying \eqref{eq:power}, where $p_{S^n}=p^n_S$ with $p_S=\mathcal{N}(0,\Lambda)$, and let $Y^n$ be the corresponding channel output through $\mathrm{AWGN}(1)$. For any decoding distribution $p_{\hat{S}^n|Y^n}$ such that the reconstruction sequence $\hat{S}^n$ follows the i.i.d. distribution $p_{\hat{S}^n}=p^n_{\hat{S}}$ with $p_{\hat{S}}=\mathcal{N}(0,\hat{\Lambda})$, we have
	\begin{align}
		\frac{1}{n}\sum\limits_{t=1}^n\mathbb{E}[\|S(t)-\hat{S}(t)\|^2]\geq D^{(u)}_c(P).\label{eq:uncoded}
	\end{align}
\end{theorem}

When $L=1$, we have
\begin{align}
 D_c^{(u)}(P)=\underline{D}_c(P)=-2\sqrt{\frac{P}{P+1}\lambda_1\hat{\lambda}_1}+\lambda_1+\hat{\lambda}_1,
 \end{align}
 which implies
\begin{align}
	\overline{D}_c(P)=-2\sqrt{\frac{P}{P+1}\lambda_1\hat{\lambda}_1}+\lambda_1+\hat{\lambda}_1.
\end{align}
In contrast, when $L=1$,
\begin{align}
	D^{(s)}_c(P)=-\frac{2P}{P+1}\sqrt{\lambda_1\hat{\lambda}_1}+\lambda_1+\hat{\lambda}_1,
\end{align}
which is strictly greater than $\overline{D}_c(P)$ for $P>0$.

When $L\geq 2$, the separation-based scheme and the uncoded scheme can be integrated into a hybrid scheme via superposition. Specifically, the encoder allocates a fraction $1-\delta$  of the power to transmit  $S^n_1$ using the uncoded scheme, referred to as the analog part, and the remaining fraction $\delta$  to transmit $S^n_2,S^n_3,\ldots,S^n_L$ using the separation-based scheme, referred to as the digital part. The decoder first decodes the digital part by treating the analog part as noise and uses it to generate $\hat{S}^n_2,\hat{S}^n_3,\ldots,\hat{S}^n_L$. It  then subtracts the digital part from the channel output and scales the residual signal to produce $\hat{S}^n_1$. The distortion associated with the analog part is
\begin{align}
	-2\sqrt{\frac{(1-\delta)P}{(1-\delta)P+1}\lambda_1\hat{\lambda}_1}+\lambda_1+\hat{\lambda}_1
\end{align}
while the distortion associated with the digital part is
\begin{align}
-2\sum\limits_{\ell=2}^L\left(\sqrt{\lambda_{\ell}\hat{\lambda}_{\ell}}-\beta(\delta)\right)_++\sum\limits_{\ell=2}^L(\lambda_{\ell}+\hat{\lambda}_{\ell}),
\end{align}
with $\beta(\delta)$ being the unique number in $(0,\sqrt{\lambda_2\hat{\lambda}_2}]$ satisfying
\begin{align}
	\prod\limits_{\ell=2}^L\max\left\{\frac{\sqrt{\lambda_{\ell}\hat{\lambda}_{\ell}}}{\beta(\delta)},1\right\}=\frac{P+1}{(1-\delta)P+1}.\label{eq:beta_delta}
\end{align}
By summing these two distortions and optimizing over the power allocation parameter $\delta$, we obtain the minimum achievable distortion under the hybrid scheme:
\begin{align}
	D^{(h)}_c(P):=&\min\limits_{\delta\in[0,1]}\left\{-2\sqrt{\frac{(1-\delta)P}{(1-\delta)P+1}\lambda_1\hat{\lambda}_1}\right.\nonumber\\
	&\hspace{-0.25in}\left.-2\sum\limits_{\ell=2}^L\left(\sqrt{\lambda_{\ell}\hat{\lambda}_{\ell}}-\beta(\delta)\right)_+\right\}+\sum\limits_{\ell=1}^L(\lambda_{\ell}+\hat{\lambda}_{\ell}).\label{eq:Dhybrid}
\end{align}

The following result shows that with an optimized $\delta$, the hybrid scheme  strictly outperforms the separation-based scheme when $P>0$, but reduces to the uncoded scheme when $P$ is sufficiently small. Its proof can be found in Appendix \ref{app:comparison}.

\begin{theorem}\label{thm:comparison}
	For $P>0$,
	\begin{align}
		D^{(h)}_c(P)<D^{(s)}_c(P).\label{eq:statement1}
	\end{align}
Moreover, when $L\geq 2$,
\begin{align}
	D^{(h)}_c(P)=D^{(u)}_c(P)\label{eq:statement2}
\end{align}
if and only if $P\in[0,P^*]$,
where
\begin{align}
P^*:=\frac{-1+\sqrt{1+\frac{\lambda_1\hat{\lambda}_1}{\lambda_2\hat{\lambda}_2}}}{2}.\label{eq:P*}
\end{align}
\end{theorem}

\begin{figure}[htbp]
	\centerline{\includegraphics[width=7cm]{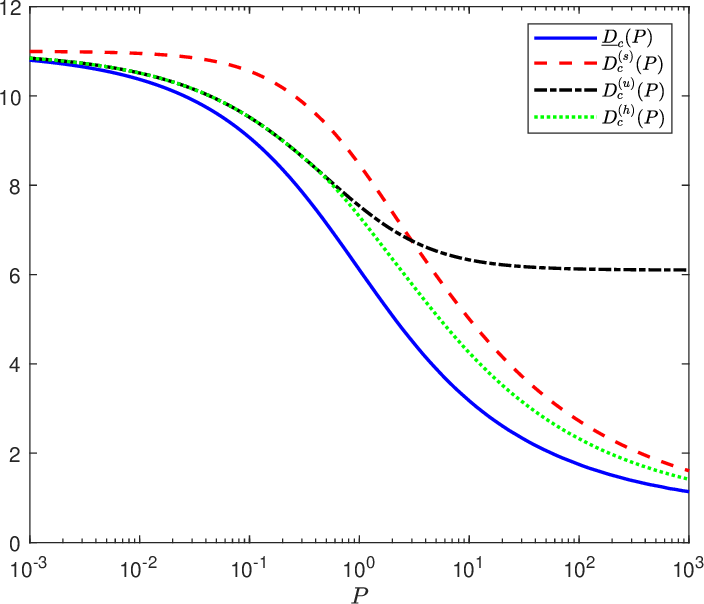}} \caption{Plots of $\underline{D}_c(P)$, $D^{(s)}_c(P)$, $D^{(u)}_c(P)$, and $D^{(h)}_c(P)$ for the case where $(\lambda_1,\lambda_2,\lambda_3)=(2,3,1)$ and $(\hat{\lambda}_1,\hat{\lambda}_2,\hat{\lambda}_3)=(3,1,1)$.}
	\label{fig:channel} 
\end{figure}

Fig. \ref{fig:channel} illustrates $\underline{D}_c(P)$, $D^{(s)}_c(P)$, $D^{(u)}_c(P)$, and $D^{(h)}_c(P)$ for a representative example. Notably,
 $\underline{D}_c(P)$, $D^{(s)}_c(P)$,  and $D^{(h)}_c(P)$ all converge to $D_{\max}$ as $P\rightarrow 0$ and to $D_{\min}$ as $P\rightarrow \infty$. While $D^{(u)}_c(P)$ follows a similar trend in the low-power regime, it saturates at $(\sqrt{\lambda}_1-\sqrt{\hat{\lambda}_1})^2+\sum_{\ell=2}^L(\lambda_{\ell}+\hat{\lambda}_{\ell})$ in the high-power limit. This occurs because the uncoded scheme transmits only the first component of each source symbol, preventing it from utilizing additional power to reduce the distortion with respect to the remaining components. It can also be seen that $D^{(h)}_c(P)$ is strictly below $D^{(s)}_c(P)$ for $P>0$ and coincides with $D^{(u)}_c(P)$ for sufficiently small $P$. On the other hand,  $D^{(h)}_c(P)$ still falls short of matching $\underline{D}_c(P)$, the minimum  distortion achievable with unlimited common randomness, whenever $P>0$. The exact characterization of $\overline{D}_c(P)$ remains unknown, though it must lie somewhere between $D^{(h)}_c(P)$ and $\underline{D}_c(P)$. 



\section{Conclusion}\label{sec:conclusion}

We have studied the problem of Gaussian Wasserstein optimal transport with commutative covariance matrices under rate, dimension, and channel constraints. The extension beyond commutative covariance matrices requires more advanced analytical techniques, which will be addressed in  future work. Notably, the Gaussian distribution represents the worst-case scenario under the squared error distortion measure. Therefore, our findings can serve as a useful reference point for understanding the limits and behavior of transport problems in more general settings.

 Overall, constrained optimal transport provides a unified framework with broad applicability across machine learning, information theory, and signal processing, opening up several promising research avenues. In this regard, our work makes an initial attempt to explore these possibilities, and we hope it will inspire further studies that delve deeper into the nuances of this theoretical framework and its real-world applications.

\appendices

\section{Proof of Theorem \ref{thm:ratecr} }\label{app:ratecr}

In light of \cite[Section III.A]{SLY15J2}, 
\begin{align}
	\underline{D}_r(R)=&\inf\limits_{p_{S\hat{S}}\in\Pi(\mathcal{N}(0,\Lambda),\mathcal{N}(0,\hat{\Lambda}))}\mathbb{E}[\|S-\hat{S}\|^2]\\
	&\mbox{s.t.}\quad I(S;\hat{S})\leq R.
\end{align}
For $p_{S\hat{S}}\in\Pi(\mathcal{N}(0,\Lambda),\mathcal{N}(0,\hat{\Lambda}))$, we have 
\begin{align}
	I(S;\hat{S})&=h(S)+h(\hat{S})-h(S,\hat{S})\nonumber\\
	&\geq h(S)+h(\hat{S})-\sum\limits_{\ell=1}^Lh(S_{\ell},\hat{S}_{\ell})\nonumber\\
	&=\sum\limits_{\ell=1}^Lh(S_{\ell})+\sum\limits_{\ell=1}^Lh(\hat{S}_{\ell})-\sum\limits_{\ell=1}^Lh(S_{\ell},\hat{S}_{\ell})\nonumber\\
	&=\sum\limits_{\ell=1}^LR_{\ell},
	\end{align}
and
\begin{align}
	\mathbb{E}[\|S-\hat{S}\|^2]&=\sum\limits_{\ell=1}^L\left(\lambda_{\ell}+\hat{\lambda}_{\ell}-2\rho_{\ell}\sqrt{\lambda_{\ell}\hat{\lambda}_{\ell}}\right),
\end{align}
where $R_{\ell}:=I(S_{\ell};\hat{S}_{\ell})$, and $\rho_{\ell}$ denotes the correlation coefficient of $S_{\ell}$ and $\hat{S}_{\ell}$ for $\ell=1,2,\ldots,L$. 
Note that
\begin{align}
	R_{\ell}\geq\frac{1}{2}\log\left(\frac{1}{1-\rho^2_{\ell}}\right),
\end{align}
which implies
\begin{align}
	\rho_{\ell}\leq\sqrt{1-2^{-2R_{\ell}}},\quad\ell=1,2,\ldots,L.
\end{align}
Therefore, $\underline{D}_r(R)$ is bounded below by the solution to the following convex optimization problem:
\begin{align}
	&\hspace{-0.1in}\min\limits_{(R_1,R_2\ldots,R_L)\in\mathbb{R}^L_+}\sum\limits_{\ell=1}^L\left(\lambda_{\ell}+\hat{\lambda}_{\ell}-2\sqrt{(1-2^{-2R_{\ell}})\lambda_{\ell}\hat{\lambda}_{\ell}}\right)\\
	&\hspace{-0.1in}\mbox{s.t.}\quad \sum\limits_{\ell=1}^LR_{\ell}\leq R.\label{eq:constrainteq1}
\end{align}
Define the Lagrangian
\begin{align}
\underline{G}:=\sum\limits_{\ell=1}^L\left(\lambda_{\ell}+\hat{\lambda}_{\ell}-2\sqrt{(1-2^{-2R_{\ell}})\lambda_{\ell}\hat{\lambda}_{\ell}}\right)+\underline{\nu}\sum\limits_{\ell=1}^nR_{\ell},
\end{align}
where $\underline{\nu}\geq 0$. It can be verified that
\begin{align}
	\frac{\mathrm{d}\underline{G}}{\mathrm{d}R_{\ell}}=-\frac{(2\ln2)2^{-2R_{\ell}}}{\sqrt{1-2^{-2R_{\ell}}}}\sqrt{\lambda_{\ell}\hat{\lambda}_{\ell}}+\underline{\nu},\quad\ell=1,2,\ldots,L.
\end{align}
For $\ell=1,2,\ldots,L$, setting $\frac{\mathrm{d}\underline{G}}{\mathrm{d}R_{\ell}}=0$ gives
\begin{align}
	R_{\ell}&=\frac{1}{2}\log\left(\frac{1+\sqrt{1+\frac{(16\ln^22)}{\underline{v}^2}\lambda_{\ell}\hat{\lambda}_{\ell}}}{2}\right)\nonumber\\
	&=\frac{1}{2}\log\left(\frac{1+\sqrt{1+\alpha\lambda_{\ell}\hat{\lambda}_{\ell}}}{2}\right),
\end{align}
 where $\alpha:=\frac{(16\ln^22)}{\underline{v}^2}$.
We obtain the minimizer $(\underline{R}_1(R),\underline{R}_2(R),\ldots,\underline{R}_L(R))$, as defined in \eqref{eq:rate1}, by choosing $\alpha$ to be the unique nonnegative number that satisfies the constraint in \eqref{eq:constrainteq1} with equality. The proof is complete since this lower bound 
is attained when $S_{\ell}$ and $\hat{S}_{\ell}$ are jointly Gaussian with the correlation coefficient $\rho_{\ell}=\sqrt{1-2^{-2\underline{R}_{\ell}(R)}}$ for $\ell=1,2,\ldots,L$, and the pairs
$(S_1,\hat{S}_1), (S_2,\hat{S}_2), \ldots, (S_L,\hat{S}_L)$ are mutually independent.


\section{Proof of Theorem \ref{thm:ratencr}}\label{app:ratencr}

In light of \cite[Theorem 1]{XLCZ24}, $\overline{D}_r(R)$ is given by the solution to the following optimization problem:
\begin{align}
	&\inf\limits_{p_{U|S},p_{\hat{U}|\hat{S}}}\mathbb{E}[\|S-U\|^2]+W^2_2(p_U,p_{\hat{U}})+\mathbb{E}[\|\hat{S}-\hat{U}\|^2]\\
	&\mbox{s.t.}\quad \max\{I(S;U);I(\hat{S};\hat{U})\}\leq R,\label{eq:const1}\\
	&\hspace{0.325in}\mathbb{E}[S|U]=U\mbox{ and } \mathbb{E}[\hat{S}|\hat{U}]=\hat{U}\mbox{ almost surely},\label{eq:const2}
\end{align}
where $S\sim\mathcal{N}(0,\Lambda)$ and $\hat{S}\sim\mathcal{N}(0,\hat{\Lambda})$. 
Consider $U:=(U_1,U_2,\ldots,U_L)^T$ and $\hat{U}:=(\hat{U}_1,\hat{U}_2,\ldots,\hat{U}_L)^T$ that satisfy \eqref{eq:const1} and \eqref{eq:const2}. Let $\xi_{\ell}:=\mathbb{E}[U^2_{\ell}]$, $\hat{\xi}_{\ell}:=\mathbb{E}[\hat{U}^2_{\ell}]$, $R_{\ell}:=I(S_{\ell};U_{\ell})$, and $\hat{R}_{\ell}:=I(\hat{S}_{\ell};\hat{U}_{\ell})$ for $\ell=1,2,\ldots,L$.
We have 
\begin{align}
	&\mathbb{E}[\|S-U\|^2]=\sum\limits_{\ell=1}^L(\lambda_{\ell}-\xi_{\ell}),\label{eq:aaa1}\\
	&\mathbb{E}[\|\hat{S}-\hat{U}\|^2]=\sum\limits_{\ell=1}^L(\hat{\lambda}_{\ell}-\hat{\xi}_{\ell}),\label{eq:aaa2}\\
	&W^2_2(p_U,p_{\hat{U}})\geq\sum\limits_{\ell=1}^L(\sqrt{\xi_{\ell}}-\sqrt{\hat{\xi}}_{\ell})^2.\label{eq:aaa3}
\end{align}
Moreover,
\begin{align}
	&I(S;U)\geq\sum\limits_{\ell=1}^LR_{\ell},\\
	&I(\hat{S};\hat{U})\geq\sum\limits_{\ell=1}^L\hat{R}_{\ell}.
\end{align}
It can be verified that for $\ell=1,2,\ldots,L$,
\begin{align}
	&\xi_{\ell}\leq(1-2^{-2R_{\ell}})\lambda_{\ell},\\
	&\hat{\xi}_{\ell}\leq(1-2^{-2\hat{R}_{\ell}})\hat{\lambda}_{\ell}.
\end{align}
Therefore, $\overline{D}_r(R)$ is bounded below by the solution to the following optimization problem:
\begin{align}
	&\hspace{-0.1in}\min\limits_{(R_1,R_2,\ldots,R_L),(\hat{R}_1,\hat{R}_2,\ldots,\hat{R}_L)\in\mathbb{R}^L_+}\sum\limits_{\ell=1}^L\left(\lambda_{\ell}+\hat{\lambda}_{\ell}-2\sqrt{(1-2^{-2R_{\ell}})}\right.\nonumber\\
	&\hspace{1.7in}\left.\times \sqrt{(1-2^{-2\hat{R}_{\ell}})\lambda_{\ell}\hat{\lambda}_{\ell}}\right)\\
	&\hspace{-0.1in}\mbox{s.t.}\quad \max\left\{\sum\limits_{\ell=1}^LR_{\ell},\sum\limits_{\ell=1}^L\hat{R}_{\ell}\right\}\leq R.
\end{align}
In view of the fact that
\begin{align}
	(1-2^{-2R_{\ell}})(1-2^{-2\hat{R}_{\ell}})
	\leq (1-2^{-(R_{\ell}+\hat{R}_{\ell})})^2,
\end{align}
there is no loss of generality in assuming $R_{\ell}=\hat{R}_{\ell}$ 
for $\ell=1,2,\ldots,L$. So the optimization problem reduces to
\begin{align}
	&\min\limits_{(R_{1},R_2,\ldots,R_{\ell})\in\mathbb{R}^L_+}\sum\limits_{\ell=1}^L\left(\lambda_{\ell}+\hat{\lambda}_{\ell}-2(1-2^{-2R_{\ell}})\sqrt{\lambda_{\ell}\hat{\lambda}_{\ell}}\right)\\
	&\mbox{s.t.}\quad\sum\limits_{\ell=1}^LR_{\ell}\leq R.\label{eq:constraintequality2}
\end{align}
Define the Lagrangian
\begin{align}
	\overline{G}:=\sum\limits_{\ell=1}^L\left(\lambda_{\ell}+\hat{\lambda}_{\ell}-2(1-2^{-2R_{\ell}})\sqrt{\lambda_{\ell}\hat{\lambda}_{\ell}}\right)+\overline{\nu}\sum\limits_{\ell=1}^nR_{\ell},
\end{align}
where $\overline{\nu}\geq 0$. Note that
\begin{align}
	\frac{\mathrm{d}\overline{G}}{\mathrm{d}R_{\ell}}=-(4\ln 2)2^{-2R_{\ell}}\sqrt{\lambda_{\ell}\hat{\lambda}_{\ell}}+\overline{\nu},\quad\ell=1,2,\ldots,L.
\end{align}
For $\ell=1,2,\ldots,L$, setting $\frac{\mathrm{d}\overline{G}}{\mathrm{d}R_{\ell}}=0$ and taking into account the constraint $R_{\ell}\geq 0$  gives 
\begin{align}
	R_{\ell}&=\frac{1}{2}\log^+\left(\frac{(4\ln2)\sqrt{\lambda_{\ell}\hat{\lambda}_{\ell}}}{\overline{\nu}}\right)\nonumber\\
	&=\frac{1}{2}\log^+\left(\frac{\sqrt{\lambda_{\ell}\hat{\lambda}_{\ell}}}{\beta}\right),
\end{align}
where $\beta:=\frac{\overline{\nu}}{4\ln2}$. We obtain the minimizer  $(\overline{R}_1(R),\overline{R}_2(R),\ldots,\overline{R}_L(R))$, as defined in \eqref{eq:rate2}, by choosing $\beta$ to be the unique number in $(0,\sqrt{\lambda_1\hat{\lambda}_1}]$ that satisfies the constraint in \eqref{eq:constraintequality2} with equality.
The proof is complete since 
this lower bound is attained when
\begin{enumerate}
	\item $U$ is jointly Gaussian with $S$ such that the pairs $(S_1,U_1),(S_2,U_2),\ldots,(S_L,U_L)$ are mutually independent, and for $\ell=1,2,\ldots,L$, the covariance matrix of $(S_{\ell},U_{\ell})$ is 
	\begin{align}
		\left(\begin{matrix}
			\lambda_{\ell} & (1-2^{-2\overline{R}_{\ell}(R)})\lambda_{\ell}\\
			(1-2^{-2\overline{R}_{\ell}(R)})\lambda_{\ell}& (1-2^{-2\overline{R}_{\ell}(R)})\lambda_{\ell}
		\end{matrix}\right),
	\end{align}

	\item $\hat{U}$ is jointly Gaussian with $\hat{S}$ such that the pairs $(\hat{S}_1,\hat{U}_1),(\hat{S}_2,\hat{U}_2),\ldots,(\hat{S}_L,\hat{U}_L)$ are mutually independent, and for $\ell=1,2,\ldots,L$, the covariance matrix of $(\hat{S}_{\ell},\hat{U}_{\ell})$ is 
	\begin{align}
		\left(\begin{matrix}
			\hat{\lambda}_{\ell} & (1-2^{-2\overline{R}_{\ell}(R)})\hat{\lambda}_{\ell}\\
			(1-2^{-2\overline{R}_{\ell}(R)})\hat{\lambda}_{\ell}& (1-2^{-2\overline{R}_{\ell}(R)})\hat{\lambda}_{\ell}
		\end{matrix}\right).
	\end{align}	
\end{enumerate}


It is instructive to compare $\overline{D}_r(R)$ with the greedy solution proposed in \cite{LZCK22,LZCK22J}.
Let $p_{U'|S}$ and $p_{\hat{U}'|\hat{S}}$ be the minimizers of 
\begin{align}
	&\min\limits_{p_{U|S}}\mathbb{E}[\|S-U\|^2]\\
	&\mbox{s.t.}\quad I(S;U)\leq R,
\end{align}
and
\begin{align}
	&\min\limits_{p_{\hat{U}|S}}\mathbb{E}[\|\hat{S}-\hat{U}\|^2]\\
	&\mbox{s.t.}\quad I(\hat{S};\hat{U})\leq R,
\end{align}
respectively.
According to the classical reverse waterfilling formula \cite[Theorem 13.3.3]{CT91}, we have
\begin{align}
	\mathbb{E}[\|S-U'\|^2]=\sum\limits_{\ell=1}^L2^{-2R'_{\ell}(R)}\lambda_{\ell},\label{eq:g1}
\end{align}
where
\begin{align}
	R'_{\ell}(R):=
		\frac{1}{2}\log^+\left(\frac{\lambda_{\ell}}{\varrho}\right),\quad\ell=1,2,\ldots,L,
\end{align}
with $\varrho$ being the unique number in $(0,\max\{\lambda_1,\lambda_2,\ldots,\lambda_L\}]$ satisfying 
\begin{align}
	\frac{1}{2}\sum\limits_{\ell=1}^L\log^+\left(\frac{\lambda_{\ell}}{\varrho}\right)=R.
\end{align}
Similary, 
\begin{align}
	\mathbb{E}[\|\hat{S}-\hat{U}'\|^2]=\sum\limits_{\ell=1}^L2^{-2\hat{R}'_{\ell}(R)}\hat{\lambda}_{\ell},\label{eq:g2}
\end{align}
where
\begin{align}
	\hat{R}'_{\ell}(R):=
		\frac{1}{2}\log^+\left(\frac{\hat{\lambda}_{\ell}}{\hat{\varrho}}\right),\quad\ell=1,2,\ldots,L,
\end{align}
with $\hat{\varrho}$ being the unique number in $(0,\max\{\hat{\lambda}_1,\hat{\lambda}_2,\ldots,\hat{\lambda}_L\}]$ satisfying 
\begin{align}
	\frac{1}{2}\sum\limits_{\ell=1}^L\log^+\left(\frac{\hat{\lambda}_{\ell}}{\hat{\varrho}}\right)=R.
\end{align}
Moreover, it can be verified that
\begin{align}
	&\hspace{-0.1in}W^2_2(p_{U'},p_{\hat{U}'})\nonumber\\
	&\hspace{-0.1in}=\sum\limits_{\ell=1}^L\left(\sqrt{(1-2^{-2R'_{\ell}(R)})\lambda_{\ell}}-\sqrt{(1-2^{-2\hat{R}'_{\ell}(R)})\hat{\lambda}_{\ell}}\right)^2.\label{eq:g3}
\end{align}
Since $U$ and $U'$ automatically satisfy \eqref{eq:const2}, summing \eqref{eq:g1}, \eqref{eq:g2}, and \eqref{eq:g3} yields the following upper bound on $\overline{D}_r(R)$:
\begin{align}
	&\sum\limits_{\ell=1}^L\left(\lambda_{\ell}+\hat{\lambda}_{\ell}-2\sqrt{(1-2^{-2R'_{\ell}(R)})(1-2^{-2\hat{R}'_{\ell}(R)})\lambda_{\ell}\hat{\lambda}_{\ell}}\right)\nonumber\\
	&=:\overline{D}'_r(R)
\end{align}
This upper bound is not tight, except in certain special cases (e.g., $L=1$ or $\Lambda=\hat{\Lambda}$). Therefore, blindly applying the  reverse-waterfilling-based  quantiztion and dequantization strategies is in general suboptimal for rate-constrained optimal transport.

\begin{figure}[htbp]
	\centerline{\includegraphics[width=7cm]{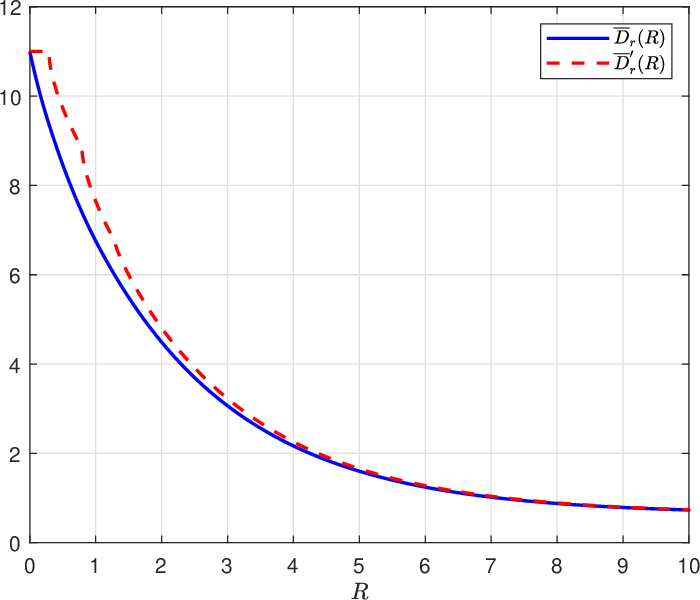}} \caption{Plots of $\overline{D}_r(R)$ and $\overline{D}'_r(R)$ for the case where $(\lambda_1,\lambda_2,\lambda_3)=(2,3,1)$ and $(\hat{\lambda}_1,\hat{\lambda}_2,\hat{\lambda}_3)=(3,1,1)$.}
	\label{fig:greedy} 
\end{figure}

Fig. \ref{fig:greedy} compares $\overline{D}_r(R)$ and $\overline{D}'_r(R)$ for an illustrative example. It can be seen that $\overline{D}'_r(R)$ is indeed suboptimal. In particular, $\overline{D}'_r(R)=D_{\max}$ when $R$ is sufficiently close to zero. This occurs because the index sets corresponding to the largest $\lambda_{\ell}$ and the largest $\hat{\lambda}_{\ell}$ are disjoint, leading to the undesirable situation in the low-rate regime where $R'_{\ell}(R)\hat{R}'_{\ell}(R)=0$ for all $\ell$. In contrast, for $\overline{D}_r(R)$, the rates allocated to $S_{\ell}$ and $\hat{S}_{\ell}$ are both given by $\overline{R}_{\ell}(R)$, 
effectively avoiding this issue.

\section{Proof of Theorem \ref{thm:dimension}}\label{app:dimension}


There is no loss of generality in assuming $K\in\{1,2,\ldots,L\}$. Let $Z:=\mathbb{E}[S|\phi(S)]$. It can be verified that
\begin{align}
	&\mathbb{E}[\|S-\hat{S}\|^2]\nonumber\\
	&=\mathbb{E}[\|S-Z\|^2]+\mathbb{E}[\|Z-\hat{S}\|^2]\nonumber\\
	&\geq\mathbb{E}[\|S-Z\|^2]+W^2_2(\mathcal{N}(0,\Delta),\mathcal{N}(0,\hat{\Lambda}))\nonumber\\
	&=\mathrm{tr}(\Lambda-\Delta)+\mathrm{tr}\left(\Delta+\hat{\Lambda}-2(\hat{\Lambda}^{\frac{1}{2}}\Delta\hat{\Lambda}^{\frac{1}{2}})^{\frac{1}{2}}\right)\nonumber\\
	&=\mathrm{tr}\left(\Lambda+\hat{\Lambda}-2(\hat{\Lambda}^{\frac{1}{2}}\Delta\hat{\Lambda}^{\frac{1}{2}})^{\frac{1}{2}}\right),
\end{align}
where $\Delta$ denotes the covariance matrix of $Z$. Clearly, $0\preceq\Delta\preceq\Lambda$ and $\mathrm{rank}(\Delta)\leq K$. Therefore, $\mathbb{E}[\|S-\hat{S}\|^2]$ is bounded below by  the solution to the following optimization problem:
\begin{align}
	&\min\limits_{\Delta}\mathrm{tr}\left(\Lambda+\hat{\Lambda}-2(\hat{\Lambda}^{\frac{1}{2}}\Delta\hat{\Lambda}^{\frac{1}{2}})^{\frac{1}{2}}\right)\label{eq:objectiveone}\\
	&\mbox{s.t.}\quad 0\preceq\Delta\preceq\Lambda,\quad \mathrm{rank}(\Delta)\leq K.\label{eq:constraintone}
\end{align}
Let $\Xi:=(\hat{\Lambda}^{\frac{1}{2}}\Delta\hat{\Lambda}^{\frac{1}{2}})^{\frac{1}{2}}$. Since
the square root is operator monotone, it follows by the L\"{o}wner-Heinz theorem that
$\Xi\preceq(\hat{\Lambda}^{\frac{1}{2}}\Lambda\hat{\Lambda}^{\frac{1}{2}})^{\frac{1}{2}}=\Lambda^{\frac{1}{2}}\hat{\Lambda}^{\frac{1}{2}}$. Moreover, $\mathrm{rank}(\Xi)=\mathrm{rank}(\Delta)$.
As a consequence, 
 we can establish a lower bound on $\mathbb{E}[\|S-\hat{S}\|^2]$ by relaxing the optimization problem in \eqref{eq:objectiveone}--\eqref{eq:constraintone} to
\begin{align}
	&\min\limits_{\Xi}\mathrm{tr}\left(\Lambda+\hat{\Lambda}-2\Xi\right)\label{eq:objective3}\\
	&\mbox{s.t.}\quad 0\preceq\Xi\preceq\Lambda^{\frac{1}{2}}\hat{\Lambda}^{\frac{1}{2}},\quad \mathrm{rank}(\Xi)\leq K.\label{eq:constraint}
\end{align}
Note that \eqref{eq:constraint} implies 
\begin{align}
	\mathrm{tr}(\Xi)\leq\sum\limits_{\ell=1}^K\sqrt{\lambda_{\ell}\hat{\lambda}_{\ell}}
\end{align}
according to \cite[Corollary 7.7.4]{HJ85}. This leads to the desired lower bound:
\begin{align}
	&\mathbb{E}[\|S-\hat{S}\|^2]\nonumber\\&\geq\mathrm{tr}(\Lambda+\hat{\Lambda})-2\sum\limits_{\ell=1}^K\sqrt{\lambda_{\ell}\hat{\lambda}_{\ell}}\nonumber\\
	&=\sum\limits_{\ell=1}^K\left(\sqrt{\lambda_{\ell}}-\sqrt{\hat{\lambda}_{\ell}}\right)^2+\sum\limits_{\ell=K+1}^L(\lambda_{\ell}+\hat{\lambda}_{\ell}),
\end{align}
which can be achieved by setting the first $K$ components of $\hat{S}$ to be the scaled versions of the corresponding components of $S$ and generating the remaining components of $\hat{S}$ from scratch.



\section{Proof of Theorem \ref{thm:linear}}\label{app:linear}

It is more convenient to represent $S^n$ and $\hat{S}^n$ as two $nL$-dimensional Gaussian random vectors with covariance matrices 
\begin{align*}
	\Lambda_n:=\left(\begin{matrix}
		\Lambda & 0 & \cdots & 0\\
		0 & \Lambda & \cdots & 0\\
		\vdots & \vdots & \ddots & \vdots\\
		0 & 0 & \cdots & \Lambda
	\end{matrix}\right), \quad \hat{\Lambda}_n:=\left(\begin{matrix}
	\hat{\Lambda} & 0 & \cdots & 0\\
	0 & \hat{\Lambda} & \cdots & 0\\
	\vdots & \vdots & \ddots & \vdots\\
	0 & 0 & \cdots & \hat{\Lambda}
\end{matrix}\right),
\end{align*}
respectively. Since $\phi$ is linear, we can write $X^n=A_nS^n$, where $A_n$ is an $n\times(nL)$ matrix. Note that the input power constraint \eqref{eq:power} is equivalent to 
\begin{align}
	\mathrm{tr}(A_n\Lambda_nA^T_n)\leq nP.\label{eq:power2}
\end{align}
Let $Z^n:=\mathbb{E}[S^n|Y^n]$. The covariance matrix of $Z^n$ is given by
\begin{align}
	\Delta_n:=\Lambda_nA^T_n(A_n\Lambda_nA^T_n+I)^{-1}A_n\Lambda_n.
\end{align}
It can be verified that
\begin{align}
	&\mathbb{E}[\|S^n-\hat{S}^n\|^2]\nonumber\\
	&=\mathbb{E}[\|S^n-Z^n\|^2]+\mathbb{E}[\|Z^n-\hat{S}^n\|^2]\nonumber\\
	&\geq\mathbb{E}[\|S^n-Z^n\|^2]+W^2_2(\mathcal{N}(0,\Delta_n),\mathcal{N}(0,\hat{\Lambda}_n))\nonumber\\
	&=\mathrm{tr}(\Lambda_n-\Delta_n)+\mathrm{tr}\left(\Delta_n+\hat{\Lambda}_n-2(\hat{\Lambda}_n^{\frac{1}{2}}\Delta_n\hat{\Lambda}_n^{\frac{1}{2}})^{\frac{1}{2}}\right)\nonumber\\
	&=\mathrm{tr}\left(\Lambda_n+\hat{\Lambda}_n-2(\hat{\Lambda}_n^{\frac{1}{2}}\Delta_n\hat{\Lambda}_n^{\frac{1}{2}})^{\frac{1}{2}}\right)\nonumber\\
	&=-2\mathrm{tr}\left((\hat{\Lambda}_n^{\frac{1}{2}}\Delta_n\hat{\Lambda}_n^{\frac{1}{2}})^{\frac{1}{2}}\right)+n\sum\limits_{\ell=1}^L(\lambda_{\ell}+\hat{\lambda}_{\ell}).
\end{align}
Let $\sigma_i(M)$ denote the $i$-th largest eigenvalue of matrix $M$. We have
\begin{align}
\mathrm{tr}\left((\hat{\Lambda}_n^{\frac{1}{2}}\Delta_n\hat{\Lambda}_n^{\frac{1}{2}})^{\frac{1}{2}}\right)=\sum\limits_{i=1}^{nL}\sqrt{\sigma_i(\hat{\Lambda}_n^{\frac{1}{2}}\Delta_n\hat{\Lambda}_n^{\frac{1}{2}})}.	
\end{align}
Since $\mathrm{rank}(\Delta_n)\leq n$, it follows that $\sigma_i(\hat{\Lambda}_n^{\frac{1}{2}}\Delta_n\hat{\Lambda}_n^{\frac{1}{2}})=0$ for $i>n$. As a consequence, 
\begin{align}
	\mathrm{tr}\left((\hat{\Lambda}_n^{\frac{1}{2}}\Delta_n\hat{\Lambda}_n^{\frac{1}{2}})^{\frac{1}{2}}\right)=\sum\limits_{i=1}^{n}\sqrt{\sigma_i(\hat{\Lambda}_n^{\frac{1}{2}}\Delta_n\hat{\Lambda}_n^{\frac{1}{2}})}.	
\end{align}
For $i=1,2,\ldots,n$,
\begin{align}
	&\sigma_i(\hat{\Lambda}_n^{\frac{1}{2}}\Delta_n\hat{\Lambda}_n^{\frac{1}{2}})\nonumber\\&=\sigma_i(\hat{\Lambda}_n\Delta_n)\nonumber\\
	&=\sigma_i(\hat{\Lambda}_n\Lambda_nA^T_n(A_n\Lambda_nA^T_n+I)^{-1}A_n\Lambda_n)\nonumber\\
	&=\sigma_i(\Lambda^{\frac{1}{2}}_n\hat{\Lambda}_n\Lambda_nA^T_n(A_n\Lambda_nA^T_n+I)^{-1}A_n\Lambda^{\frac{1}{2}}_n)\nonumber\\
	&\leq\sigma_1(\Lambda^{\frac{1}{2}}_n\hat{\Lambda}_n\Lambda^{\frac{1}{2}}_n)\sigma_i(\Lambda^{\frac{1}{2}}_nA^T_n(A_n\Lambda_nA^T_n+I)^{-1}A_n\Lambda^{\frac{1}{2}}_n)\nonumber\\
	&=\lambda_1\hat{\lambda}_1\sigma_i(\Lambda^{\frac{1}{2}}_nA^T_n(A_n\Lambda_nA^T_n+I)^{-1}A_n\Lambda^{\frac{1}{2}}_n)\nonumber\\
		&=\lambda_1\hat{\lambda}_1\sigma_i(A_n\Lambda_nA^T_n(A_n\Lambda_nA^T_n+I)^{-1})\nonumber\\
	&=\frac{\sigma_i(A_n\Lambda_nA^T_n)}{\sigma_i(A_n\Lambda_nA^T_n)+1}\lambda_1\hat{\lambda}_1,
\end{align}
where the inequality follows by \cite[Corollary 4.6.3]{WWJ06}.
Moreover, \eqref{eq:power2} can be written equivalently as
\begin{align}
	\sum\limits_{i=1}^n\sigma_i(A_n\Lambda_nA^T_n)\leq nP.
\end{align}
Therefore, we have
\begin{align}
	\mathbb{E}[\|S^n-\hat{S}^n\|^2]\geq -2\zeta+n\sum\limits_{\ell=1}^L(\lambda_{\ell}+\hat{\lambda}_{\ell}),\label{eq:subzeta}
\end{align}
where
\begin{align}
\zeta:=&\max\limits_{(\sigma_1,\sigma_2,\ldots,\sigma_n)\in\mathbb{R}^n_+}\sum\limits_{i=1}^n\sqrt{\frac{\sigma_i}{\sigma_i+1}\lambda_1\hat{\lambda}_1}\label{eq:maxsigma}\\
&\mbox{s.t.}\quad\sum\limits_{i=1}^n\sigma_i\leq nP.
	\end{align}
Since $\sqrt{\frac{\sigma}{\sigma+1}}$ is concave in $\sigma$ for $\sigma\geq0$, the maximum in \eqref{eq:maxsigma} is attained at $\sigma_1=\sigma_2=\ldots=\sigma_n=P$, and consequently,
\begin{align}
	\zeta=n\sqrt{\frac{P}{P+1}\lambda_1\hat{\lambda}_1}.\label{eq:zeta}
\end{align}
Substituting \eqref{eq:zeta} into \eqref{eq:subzeta} and dividing both sides by $n$ yields the desired lower bound.

\section{Proof of Theorem \ref{thm:comparison}}\label{app:comparison}

It suffices to consider the case $L\geq 2$. In view of \eqref{eq:Dsep} and Theorem \ref{thm:ratencr},
\begin{align}
D^{(s)}_c(P)=-2\sum\limits_{\ell=1}^L\left(\sqrt{\lambda_{\ell}\hat{\lambda}_{\ell}}-\beta\right)_++\sum\limits_{\ell=1}^L(\lambda_{\ell}+\hat{\lambda}_{\ell}),
\end{align}
with $\beta$ being the unique number in $(0,\sqrt{\lambda_1\hat{\lambda}_1}]$ satisfying
\begin{align}
	\prod\limits_{\ell=1}^L\max\left\{\frac{\sqrt{\lambda_{\ell}\hat{\lambda}_{\ell}}}{\beta},1\right\}=P+1.
\end{align}
When $P>0$, we must have $\beta<\sqrt{\lambda_1\hat{\lambda}_1}$. Let 
\begin{align}
	\delta^*:=1-\frac{\sqrt{\lambda_1\hat{\lambda}_1}-\beta}{\beta P}.
\end{align}
It can be verified that $\delta^*\in[0,1)$ and
\begin{align}
	\prod\limits_{\ell=2}^L\max\left\{\frac{\sqrt{\lambda_{\ell}\hat{\lambda}_{\ell}}}{\beta},1\right\}=\frac{P+1}{(1-\delta^*)P+1}.\label{eq:beta_delta2}
\end{align}
In light of \eqref{eq:beta_delta}, \eqref{eq:Dhybrid}, and \eqref{eq:beta_delta2},
\begin{align}
	D^{(h)}_c(P)\leq&-2\sqrt{\frac{(1-\delta^*)P}{(1-\delta^*)P+1}\lambda_1\hat{\lambda}_1}\nonumber\\
	&-2\sum\limits_{\ell=2}^L\left(\sqrt{\lambda_{\ell}\hat{\lambda}_{\ell}}-\beta\right)_++\sum\limits_{\ell=1}^L(\lambda_{\ell}+\hat{\lambda}_{\ell}).
\end{align}
Therefore,
\begin{align}
	&D^{(s)}_c(P)-D^{(h)}_c(P)\nonumber\\
	&\geq-2\left(\sqrt{\lambda_1\hat{\lambda}_1}-\beta\right)+2\sqrt{\frac{(1-\delta^*)P}{(1-\delta^*)P+1}\lambda_1\hat{\lambda}_1}\nonumber\\
	&=-2\left(\sqrt{\lambda_1\hat{\lambda}_1}-\beta\right)+2\sqrt{\lambda_1\hat{\lambda}_1-\beta\sqrt{\lambda_1\hat{\lambda_1}}}\nonumber\\
	&>-2\left(\sqrt{\lambda_1\hat{\lambda}_1}-\beta\right)+2\sqrt{\lambda_1\hat{\lambda}_1-2\beta\sqrt{\lambda_1\hat{\lambda_1}}+\beta^2}\nonumber\\
	&=0.
\end{align}
This proves \eqref{eq:statement1}.

Note that
\begin{align}
	D^{(h)}_c(P)=\min\limits_{\delta\in[0,1]}\{-2f_1(\delta)-2f_2(\delta)\}+\sum\limits_{\ell=1}^L(\lambda_{\ell}+\hat{\lambda}_{\ell}),\label{eq:Dhybrid2}
\end{align}
where
\begin{align}
	&f_1(\delta):=\sqrt{\frac{(1-\delta)P}{(1-\delta)P+1}\lambda_1\hat{\lambda}_1},\nonumber\\
	&f_2(\delta):=\sum\limits_{\ell=2}^L\left(\sqrt{\lambda_{\ell}\hat{\lambda}_{\ell}}-\beta(\delta)\right)_+.
\end{align}
The proof of \eqref{eq:statement2} boils down to determining the condition under which the minimum in \eqref{eq:Dhybrid2} is attained at $\delta=0$. Clearly,
\begin{align}
	\frac{\mathrm{d}f_1(\delta)}{\mathrm{d}\delta}&=-\frac{1}{2}\sqrt{\frac{P}{(1-\delta)((1-\delta)P+1)^3}\lambda_1\hat{\lambda}_1}\nonumber\\
	&\leq-\frac{1}{2}\sqrt{\frac{P}{(P+1)^3}\lambda_1\hat{\lambda}_1}.
\end{align}
It can be verified that
\begin{align}
	f_2(\delta)=&-\kappa(\delta)\left(\frac{(1-\delta)P+1}{P+1}\prod\limits_{\ell=2}^{\kappa(\delta)+1}\sqrt{\lambda_{\ell}\hat{\lambda}_{\ell}}\right)^{\frac{1}{\kappa(\delta)}}
	\nonumber\\
	&+\sum\limits_{\ell=2}^{\kappa(\delta)+1}\sqrt{\lambda_{\ell}\hat{\lambda}_{\ell}},
\end{align}
where $\kappa(\delta)$ denotes the largest $\ell\in\{1,2,\ldots,L-1\}$ satisfying  $\sqrt{\lambda_{\ell+1}\hat{\lambda}_{\ell+1}}\geq\beta(\delta)$. Since $\kappa(\delta)$ is a piecewise constant function of $\delta$, we have
\begin{align}
	\frac{\mathrm{d}f_2(\delta)}{\mathrm{d}\delta}=\frac{P}{(1-\delta)P+1}\left(\frac{(1-\delta)P+1}{P+1}\prod\limits_{\ell=2}^{\kappa(\delta)+1}\sqrt{\lambda_{\ell}\hat{\lambda}_{\ell}}\right)^{\frac{1}{\kappa(\delta)}}\label{eq:expression}
\end{align}
within each interval of $\delta$ where $\kappa(\delta)$ is fixed. The expression in \eqref{eq:expression} is maximized when $\kappa(\delta)=1$, yielding
\begin{align}
\frac{\mathrm{d}f_2(\delta)}{\mathrm{d}\delta}\leq\frac{P}{P+1}\sqrt{\lambda_2\hat{\lambda}_2}.	
\end{align}
Therefore, 
\begin{align}
\frac{\mathrm{d}f_1(\delta)}{\mathrm{d}\delta}+\frac{\mathrm{d}f_2(\delta)}{\mathrm{d}\delta}\leq-\frac{1}{2}\sqrt{\frac{P}{(P+1)^3}\lambda_1\hat{\lambda}_1}+ \frac{P}{P+1}\sqrt{\lambda_2\hat{\lambda}_2}.
\end{align}
The solution to 
\begin{align}
-\frac{1}{2}\sqrt{\frac{P}{(P+1)^3}\lambda_1\hat{\lambda}_1}+ \frac{P}{P+1}\sqrt{\lambda_2\hat{\lambda}_2}=0
\end{align}
is given by $P=P^*$ defined in \eqref{eq:P*}. For $P\in[0,P^*]$,
\begin{align}
	\frac{\mathrm{d}f_1(\delta)}{\mathrm{d}\delta}+\frac{\mathrm{d}f_2(\delta)}{\mathrm{d}\delta}\leq 0,
\end{align} 
which implies that the minimum in \eqref{eq:Dhybrid2} is attained at $\delta=0$. On the other hand, for $P>P^*$, we have
\begin{align}
	&\left.\frac{\mathrm{d}f_1(\delta)}{\mathrm{d}\delta}\right|_{\delta=0}+\left.\frac{\mathrm{d}f_2(\delta)}{\mathrm{d}\delta}\right|_{\delta=0}\nonumber\\
	&=-\frac{1}{2}\sqrt{\frac{P}{(P+1)^3}\lambda_1\hat{\lambda}_1}+ \frac{P}{P+1}\sqrt{\lambda_2\hat{\lambda}_2}\nonumber\\
	&>0,
\end{align}
and consequently,  the minimum in \eqref{eq:Dhybrid2} is not attained at $\delta=0$. This completes the proof of \eqref{eq:statement2}.





\end{document}